\documentclass {article}
\usepackage{graphicx}

\begin{document}

\title{Are we living in a string-dominated universe?}

\author{Michael Petri\thanks{email: mpetri@bfs.de} \\Bundesamt f\"{u}r Strahlenschutz (BfS), Salzgitter, Germany}

\date{May 1, 2004}

\maketitle

\begin{abstract}
The so called holographic solution is a new exact solution to the
Einstein field equations. The solution describes a compact
self-gravitating object with properties very similar to a black
hole. Its entropy and temperature at infinity are proportional to
the Hawking result. Instead of an event horizon, the holographic
solution has a real spherical boundary membrane, situated roughly
two Planck distances outside of the object's gravitational radius.

The interior matter-state of the holographic solution is
singularity free. It consists out of string type matter, which is
densely packed. Each string occupies a transverse extension of
exactly one Planck area. This dense package of strings might be
the reason, why the solution does not collapse to a singularity.
The local string tension is inverse proportional to the average
string length. This purely classical result has its almost exact
correspondence in a recent result in string theory, published by
Mathur.

The holographic solution suggest, that string theory is relevant
not only on microscopic, but also on cosmological scales. The
large scale phenomena in the universe can be explained naturally
in a string context. Due to the zero active gravitational
mass-density of the strings, the Hubble constant in a string
dominated universe is related to its age by $H t =1$. The WMAP
measurements have determined $H t \approx 1.02 \pm 0.02$
experimentally. The nearly unaccelerated expansion expected in a
string dominated universe is compatible with the recent supernova
measurements. Under the assumption, that the cold dark matter
(CDM) consists out of strings, the ratio of CDM to baryonic matter
is estimated. We find $\Omega_{CDM} / \Omega_b \approx 6.45$.

Some arguments are given, which suggest that the universe might be
constructed hierarchically out of its most basic building blocks:
strings and membranes.

\end{abstract}

\maketitle

\section{\label{sec:intro}Introduction}

In \cite{petri/bh} a new spherically symmetric, exact solution of
the Einstein field equations with zero cosmological constant was
reported. The starting point leading to its discovery was to find
an alternative singularity-free description for a compact black
hole type object. The so called holographic solution, in short
"holostar", has a temperature (at infinity) and an entropy
proportional to the entropy and temperature of a black hole
\cite{petri/thermo}. Instead of an event horizon, the holostar has
a boundary membrane consisting out of tangential pressure situated
roughly two Planck distances outside its gravitational radius
\cite{petri/hol}. This real physical membrane has the same
properties as the - purely fictitious -"membrane" attributed to a
black hole by the so-called membrane paradigm \cite{Thorne/mem,
Price/Thorne/mem} ($P_\perp = 1 / (16 \pi r)$, $\rho = 0$). This
guarantees, that the dynamic action of a holostar on the exterior
space-time is practically identical to that of a black
hole.\footnote{Black holes are thought to be the most compact self
gravitating objects possible, by a vast majority of researchers.
However, despite years of research we don't have any definite
answers to our most fundamental questions, such as the microscopic
origin of the Hawking entropy, the nature of the singularities,
the question of information-loss, unitary vs. non-unitary
evolution and other related questions. This makes it necessary to
explore alternatives.} The interior matter state is non-singular
with a well defined temperature and energy-density at every
interior space-time region.

By studying the geometric properties of this new solution (for an
extensive treatment see \cite{petri/hol}) it turned out, that the
new solution might also serve as an alternative model for the
universe. Far away from the center the geodesic motion of massive
particles is virtually indistinguishable from that of a uniformly
expanding (or contracting) Friedmann Robertson-Walker (FRW)
universe. A very attractive feature of the new model is, that it
has practically no free parameters, which could be tuned to
observational facts. Therefore this model is very easily
falsifiable. Nevertheless, the solution fits almost perfectly -
within the measurement errors - to the henceforth available
experimental data:

\begin{itemize}
\item It predicts a definite relation between the total matter
density $\rho$ and the CMBR-temperature $T$, which is
experimentally verified within a few percent of error: $\rho / T^4
\simeq 2^6 \pi^3 \sqrt{3} / \hbar^4$ in units $c=G=1$.

\item It predicts a uniform Hubble-type expansion with $H t = 1$,
exactly. The recent WMAP measurements claim $H t = 1.02 \pm 0.02$.

\item It predicts a coasting (nearly unaccelerated) expansion with
nearly zero deceleration ($q \simeq 0$), which is compatible with
the supernova-data, if $H \approx 60-63 \, km s^{-1} MPc^{-1}$.

\item The Hubble constant is related to the microwave-background-temperature, predicting $H \approx 63 \, km s^{-1} MPc^{-1}$. This is quite close to the value $H = 71$ used in the concordance $\Lambda$CDM-model \cite{WMAP/cosmologicalParameters} and fits almost perfectly with other absolute measurements of $H$, which consistently yield values  in the range around $H \approx 60 \pm 10$.

\end{itemize}

The expansion in the holostar solution is accelerated, with the
proper acceleration falling off over time. The acceleration is not
due to a cosmological constant, which is exactly zero in the
holostar space-time. Rather, the proper acceleration in the
co-moving frame can be traced to the radial dependence of the
spherically symmetric gravitational potential, which falls off
with $r$ by a power law.

However, the interior matter state of the new solution is
puzzling. The interior pressure is highly anisotropic: The radial
pressure is negative and exactly equal, but opposite in sign to
the positive mass density. The two tangential pressure components
are zero. This is the equation of state of a classical string in
the radial direction. Thus the interior matter state can be
interpreted as a spherical arrangement of radially outlayed
strings, attached to a real physical membrane, which constitutes
the boundary of the object.

The string tension $\mu$ falls off with radius: $\mu = 1 / (8 \pi
r^2)$. The average length $l$ of the strings at radial position
$r$ is inverse proportional to the tension, $l = r^2 / (2 r_0)$,
so that $\mu \overline{l} = 1 / (16 \pi r_0)$. $r_0 \approx 2
r_{Pl}$ is a fundamental length, which can be shown to be slightly
less than 2 Planck lengths. This result is compatible with a very
recent result in string theory \cite{Mathur/2004}, according to
which the string tension of a large black hole type object falls
with its inverse length, so that the "black hole's" interior is
filled with strings extending up to the event horizon.

String theory is the domain of particle physicists and is
predominantly used to analyze the phenomena at the highest
conceivable energies, approaching or surpassing the Planck energy.
Why should a solution of the Einstein field equations with an
interior matter-state consisting out of strings be relevant to the
physics at low energies, such as the present state of the
universe? There are three quick answers to this question:

\begin{itemize}
\item The dualities of string-theory: If a certain string theory at
high energies is equivalent to a dual theory at low energies it is
difficult to justify the belief, that string theory should only be
relevant at high energies. A particular example for the relevance
of string theory at low energies is given in \cite{Mathur/2004}

\item The cold dark matter (CDM): There is overwhelming experimental evidence that our universe consists of a large fraction of cold dark, presumably non-baryonic, matter. Not much is known about it, beyond its very existence. Why not add another type of matter to the long list of CDM-candidates: strings.

\item The universe itself: The universe, as we see it today, exhibits
several properties which can be explained very naturally in a string context.
\end{itemize}

Some essential pieces of evidence for the last claim were already
pointed out beforehand. I will shortly explain in the following
paragraphs, how these definite predictions of the holographic
solution are related to its string nature. For a full treatment
the reader must be referred to the 150 pages of \cite{petri/hol}.
There the reader will find some other predictions which are quite
compatible with the observation, such as a baryon to photon ratio
$\eta \approx 10^{-9}$, a prediction for the low values of the
CMBR-quadrupole moment etc. . In \cite{petri/thermo, petri/asym}
an explanation for the origin of the matter-antimatter asymmetry
at high temperatures within the holostar space-time is given.

Due to the zero active gravitational mass-density of the strings a
string dominated universe expands uniformly with
$r \propto t$. This implies $H t = 1$, which is very well
fulfilled in our universe today.\footnote{$r \propto t$ is the
relation for (permanently) unaccelerated expansion. The large
scale motion of particles must be unaccelerated in a
string-dominated universe, because the active gravitational
mass-density $\rho + \sum{P_i}$, which determines the
acceleration/deceleration in any local Minkowski frame, is zero
for stringy matter.}

The prediction of nearly unaccelerated expansion in a string
dominated universe is also compatible with the recent
supernova-measurements. The luminosity-redshift relation for a
permanently zero deceleration parameter ($q=0$) is nearly
indistinguishable from the relations predicted by today's
preferred models with $\Omega_\Lambda \approx 0.65-0.75$ and
$\Omega_m \approx 0.25-0.35$, at least in the range of red-shifts
covered by the recent surveys ($z < 1.75$). However, with the
current available data the best fit $\Lambda$CDM model gives a
$\chi^2$-value which is roughly one standard deviation (of the
$\chi^2$-test!) lower than the $\chi^2$-value of the holostar
model, so that the $\Lambda$CDM model is preferred over the
holostar-model at one sigma confidence level.\footnote{One
standard deviation can hardly be regarded as a statistically
significant, unless one is willing to change ones predictions for
every third data sample.} A definite decision with respect to what
model provides the best description for the universe most likely
will be obtained, when more supernova-measurements in the high
z-range ($z>2$) are available, where both models differ
substantially in their predictions. See \cite{petri/hol} for a
detailed discussion.

The relation $H t =1$ for a permanently unaccelerated universe is
interesting from another perspective. If we take the radius of the
observable universe $r$ to lie in the range $13-18 \, Gy$, this
translates to $r \approx t \approx 10^{61}$ in Planck units. If
the universe was string-dominated with $r \propto t$ for all time,
the expansion will have started out from roughly a Planck volume
at the Planck time and Planck temperature. In contrast, the
standard FRW-model requires that the scale factor of the
observable universe was of order $10^{30}$ Planck-lengths at the
Planck-time and Planck-temperature. There appears to be no
fundamental reason why the universe has chosen such an odd number.

Uniform expansion in a string dominated universe can also explain
the nearly scale-invariant acoustic spectrum found in the CMBR
(see for example \cite{Peacock} and references therein), which was
mapped by WMAP \cite{WMAP/cosmologicalParameters} to a high degree
of accuracy. Strings could be an explanation for the recently
found deviations in the CMBR-spectrum from a purely Gaussian
distribution \cite{Gurzadyan/2003, Gurzadyan/2004, Hansen/2004,
Larson/2004, Schwarz/2004} and for the anisotropies suggested by
the analysis of the lower multipoles \cite{Ralston}. Furthermore
strings can give a quite natural explanation for the amplitude of
the density fluctuations in the CMBR in terms of the GUT-scale
$\delta \approx M_{Gut}^2 / M_{Planck}^2 \approx 10^{-5}$ (see
\cite[p. 316]{Peacock} and references therein).

Expansion in a string dominated universe has no horizon problem.
The relation $r \propto t$, which arises from the zero active
gravitational mass-density of the strings, guarantees that the
scale factor and the Hubble-distance are always proportional to
each other. Inflation is not necessary. Furthermore, a string
dominated universe, as described by the holostar solution, has no
cosmological constant problem. In the holostar solution the
cosmological constant is exactly zero.

A zero cosmological constant is attractive not only from an
esthetic point of view. It is well known, that string theory has
severe problems with a positive cosmological constant. The
apparent necessity for a positive cosmological constant in the
standard FRW-type models has led many string theorists to turn to
anthropic reasoning. This is unsatisfactory (although maybe
unavoidable in the long run). Anthropic arguments are "a
posteriori", i.e. they don't explain {\em why} the cosmological
constant has taken its particular value. The holographic solution
might provide an elegant way out. It enables us to explain the
phenomena in a model with {\em zero} cosmological constant. If the
holographic solution - or a generalization thereof - actually
turns out to be the correct description of the universe, this will
provide string theorists with invaluable experimental /
observational data from the low energy sector, which might be
helpful in the understanding of string theory at the high energy
limit. It might also provide string theorists with an incentive -
and most likely some guidance - to show, why the cosmological
constant {\em should} be close to zero in a self-consistent
unified theory of quantum gravity encompassing all known forces.

Therefore this new solution appears worthwhile to explore, from a
theoretical as well as an observational point of view.

In this paper, I will attempt to further develop the ideas and
insights reported in \cite{petri/bh, petri/hol, petri/thermo,
petri/charge}, with particular emphasis on the interpretation that
the interior matter state of the solution consists predominantly
out of low energy string-type matter bounded by a 2D membrane.

\section{\label{sec:holo}A short introduction to the holographic solution}

The holographic solution, on which the calculations presented in
this paper are based, solves the Einstein field equations with
zero cosmological constant exactly. It's metric in the usual
spherical (Schwarzschild) coordinate system ($t, r, \theta, \phi$)
and with the (+ - - - ) sign convention is given by:

\begin{equation}
ds^2 = B dt^2 - A dr^2 - r^2 d\Omega^2
\end{equation}

with

\begin{equation} \label{eq:metric}
B = \frac{1}{A} = \frac{r}{r_0} \theta(r-r_h) + (1-\frac{r_+}{r})
\theta(r-r_h)
\end{equation}

$r_h = r_+ + r_0$ is the boundary of the matter distribution, $r_+
= 2 M$ is its gravitational radius. $r_0$ is a scale parameter,
which has been shown in \cite{petri/charge} to be roughly twice
the Planck length. Throughout this paper natural units $c = G =
\hbar = 1$ will be used.

The matter fields of the solution, which can be derived from the
metric by simple differentiation (see for example
\cite{petri/bh}), are given by:

\begin{equation} \label{eq:rho}
\rho = \frac{1}{8 \pi r^2} \theta (r-r_h) = -P_r
\end{equation}

\begin{equation} \label{eq:Pt}
P_\perp = \frac{1}{16 \pi r_h} \delta(r-r_h)
\end{equation}

$\rho$ is the energy density, $P_r$ is the radial pressure,
$P_\perp$ describes the pressure in the two tangential
directions.

From equations (\ref{eq:rho}, \ref{eq:Pt}) one immediately sees,
that the interior matter state of the new solution is that of a
collection of strings, layed out radially and - in a sense -
attached to the 2D-membrane, which constitutes the boundary of the
matter-distribution.

Remarkably, this new - purely classical - solution fits quite well
with the theoretical expectations of string theory, i.e.
1-dimensional strings attached to D-branes, here: a 2D-membrane in
3D curved space. In fact one could say, that this new solution,
had it been found earlier, would have in a sense "predicted"
strings attached to membranes as one of the basic building blocks
of nature.

\section{\label{sec:strings}A determination of the string's transverse extension}

Let us explore the interior matter state in more detail.
According to equation (\ref{eq:rho}) the (positive) string tension
$\mu$ is given by:

\begin{equation}
\mu = -P_r = \frac{1} {8 \pi r^2}
\end{equation}

The tension falls off with an inverse square law. For large
$r$, such as the current radius of the universe ($r \approx
10^{61} r_{Pl}$), the energy density in the strings is very low,
yet almost exactly equal to the mean energy density of the
universe as we see it today. For these low energies we can be
quite confident that the classical field equation of general
relativity are an excellent approximation to the true unified
quantum theory of gravity.

The holostar solution allows us to determine the total number of
strings attached to the boundary membrane by a simple argument. It
is very well known, that any one string introduces a deficit angle
$\Delta \varphi$ in the flat (background) geometry proportional to
the string tension (see for example \cite[p. 313]{Peacock/book}):

\begin{equation} \label{eq:deficit:angle}
\Delta \varphi = 8 \pi \mu = \frac{1}{r^2}
\end{equation}

The holostar solution, however, describes a curved space-time. For
a large holostar the curvature at the membrane - almost - induces
a spherical topology, nearly indistinguishable from a black hole
of the same gravitational mass. Let us denote by $N_p$ the number
of strings (or rather string segments) attached to the boundary
membrane. The individual deficit angles of all strings must add up
(almost) to the solid angle of the sphere.\footnote{This simple
"summing up" of deficit angles over a (fixed!) flat background
space works, despite the {\em non-linearity} of the field
equations: The holostar solution resides in the "linearized
sector" of the field equations. In any spherically symmetric
problem a string equation of state $\rho + P_r = 0$ has the effect
to reduce the generally non-linear equations to a single linear
one order differential equation of a single variable, the time
coefficient of the metric $B$. The matter-density and principal
pressures are linear combinations of $B$ and its first and second
derivatives (the second derivative of $B$ is only required for the
tangential pressure). See \cite{petri/bh} for a somewhat more
detailed discussion.

Or stated somewhat differently: It is the {\em zero active
gravitational mass-density of the strings}, which allows us to
construct string theory in a linear perturbation expansion over a
fixed flat background geometry (at least for a spherically
symmetric problem), despite the non-linearity of the field
equations in the general case! Therefore in the opinion of this
author the major criticism that is often leveled against string
theory, that it "does not take the non-linearity of gravity
properly into account", "relies on a perturbation expansion over a
flat (Minkowski) background" or "requires some pre-geometry"
looses much of its bite.} We therefore get the condition $N_p
\Delta \varphi = 4 \pi$, from which the following important result
follows:

\begin{equation} \label{eq:Ns}
N_p = 4 \pi r^2 = A
\end{equation}

For large holostars this relation is exact to order $\Delta N_p
\approx 1$. $A$ is the proper area of the membrane measured in
Planck units. We find, that $N_p$ strings segments, laid out
radially on a flat Minkowski background space, actually induce a
curvature in this background space via the individual deficit
angles of the strings. If we lay out a large number of string
segments with the right tension in the radial direction we can
"create" a large black hole type object of arbitrary size by an
explicit construction.

Equation (\ref{eq:Ns}) tells us, that every string segment occupies a
membrane segment of exactly one Planck area. This result is a
genuine prediction of {\em classical} general relativity. There is no
other argument involved than the validity of the field equations
(with zero cosmological constant), spherical symmetry and the
argument, that the deficit angles of all strings must add up to
the solid angle of the sphere for a large black hole type object.

This result can be derived by an independent argument: For any
classical string the string tension $\mu = -P_r$ is nothing else
than the energy per unit length, i.e. $\mu = \delta E
/ \delta l$. Now consider a large holostar and imagine a thin
spherical concentric shell situated at radial position $r$ with
proper thickness $\delta l$. This thin shell will be punctured by
$N_p$ radial strings. If $\delta l$ is chosen small enough, no
strings will "end" within the shell. The energy $\delta E$ per
string segment is $\mu \delta l$, so that the total energy in the
shell is given by:

$$E = N_p \delta E = N_p \mu \delta l = \frac{N_p \delta l}{8 \pi r^2}$$

where $\mu = -P_r = 1 / (8 \pi r^2)$ from the holostar equations was used.

Let us compare this energy to the energy of the shell calculated
from the holostar solution. The proper volume of any thin
concentric shell is $\delta V = 4 \pi r^2 \delta l$. Using the
holostar-expression for the interior energy-density $\rho = 1 / (8
\pi r^2)$ the total energy in the shell can be calculated from the
product of energy-density times proper volume:

$$E = \rho \delta V = \frac{\delta l}{2}$$

Setting both energies equal we find that the number of strings
puncturing any spherical surface at radial position $r$ is given
by $N_p = 4 \pi r^2 = A$, i.e. exactly the result that was
obtained by the deficit angle argument. However, the second
derivation is more general. It refers to any interior concentric
thin shell at arbitrary radial position $r$. Therefore the string
separation in the tangential direction is {\em universal}: Any radial
string-segment in the holostar's interior has a tangential
extension of exactly one Planck area.

$$A_\perp = A_{Pl} = \frac{G \hbar}{c^3}$$

What does this result imply for the interior structure of a
compact self gravitating object, described by the holostar solution?

Loosely speaking the boundary membrane has $N_p$ string segments
attached and every string segment occupies a Planck area of the
membrane. The tangential (positive) pressure in the membrane can
be thought to be created by the transverse "wigglings" of the
strings attached to the boundary membrane.\footnote{Note, that the
membrane has no mass-energy, only tangential pressure. This is
obvious in the string-picture, because mass-energy resides in the
string's longitudinal dimension (at least for large strings): A
string-segment of zero-length has zero mass-energy. All string
segments are perpendicular to the membrane and the membrane has
zero thickness. Therefore the membrane's mass-energy must be
zero.}

Within the interior of the holostar solution the tangential
pressure components are zero. The negative radial pressure is
nothing else than the tension of the strings. The string tension -
and therefore the energy-density of the strings - grows, as we
approach the center. The transverse extension of the strings is
universal, meaning that the strings are {\em densely packed
throughout the whole interior}. If one approaches the holostar's
center, the number of radially outlayed strings puncturing any
sphere concentric to the center declines, as the proper area of
the sphere becomes smaller. The center is reached when there is
just one string segment of roughly Planck length left, filling out
roughly a Planck volume.\footnote{There might be modifications to
this statement at very high energies, as classical general
relativity is expected to break down at the Planck-energy.}

We can picture the holostar solution as the densest possible
collection of radially outlayed strings. The holostar's {\em
curved} classical space-time arises from an explicit construction,
by laying out a maximally dense package of string segments
radially on a flat Minkowski background space-time. It is clear
from this construction, that the holographic solution is the most
compact non-singular spherically symmetric solution for a
self-gravitating object.

This construction actually might be at the heart of the answer,
why a holostar does not collapse under it's own gravity to a
singularity, although its boundary lies just a few
Planck-distances outside of its gravitational radius: If we take
string-theory and it's prediction of a minimum transverse
dimension of the strings seriously, we logically have to accept
that it is exactly this minimum transverse dimension that prevents
the formation of singularities and at the same time allows us to
construct {\em arbitrarily extended} singularity free
self-gravitating objects, nearly as compact as black holes, whose
number of fundamental degrees of freedom (strings!) are
real\footnote{not just fictitious "boundary states" on a locally
undetectable surface in vacuum, whose position can only be
determined by knowing the whole space-times future - the event
horizon} and don't scale with volume, but with area.

\section{\label{sec:tension}On the relation between string length and tension}

We have already seen, that the number of string segments attached
to the holostar's boundary membrane is $N = A$. Any string segment
has two end points. In the very simple analysis in this section we
are only interested in the long strings, which end on the boundary
membrane. Any closed string in the interior is expected to shrink
to small overall size, suggesting a particle interpretation. We
will neglect the contribution of closed strings (=particles?) in
the following argument.

\begin{figure}[ht]
%\begin{minipage}[c]
\begin{center}
\includegraphics[width=12cm, bb=70 240 540 630]{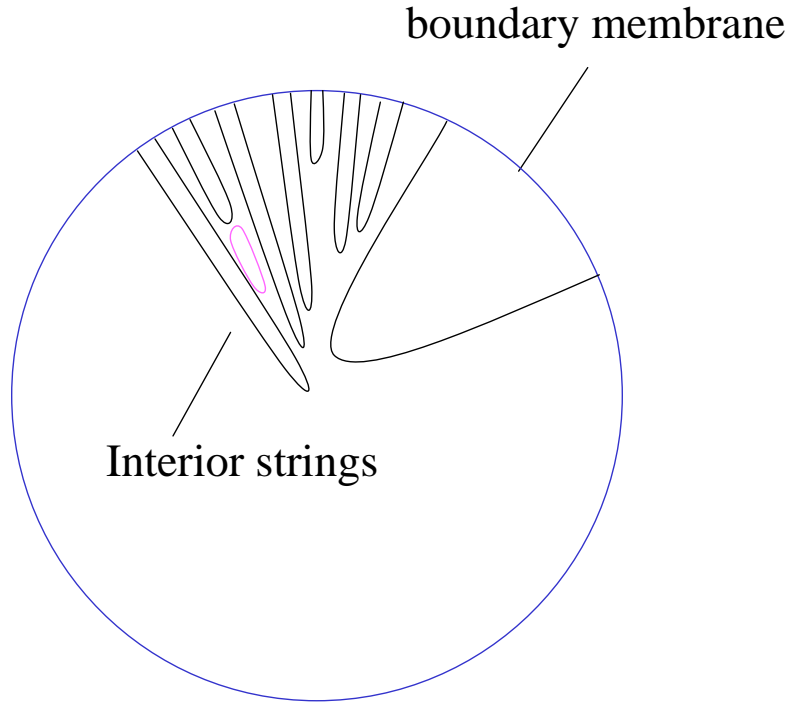}
\caption{\label{fig:strings} Arrangement of strings within the
holostar solution. Most strings will be attached to the spherical
boundary membrane. There might also be some closed loops.}
\end{center}
%\end{minipage}
\end{figure}

If there are no closed strings in the interior space-time, the two
end-points of every string must end on the boundary membrane,
which is the only structure in the holostar space-time resembling
a D-brane. According to string theory, "loose" string ends should
end on D-brands. Therefore any one string will consist out of two
segments, attached to the membrane and extending radially into the
holostar's interior. The interior string ends will join at some
radial coordinate position $r$ within the interior space-time. See
figure \ref{fig:strings} for a crude pictorial representation.

Therefore the total number of strings $N_s$ is half the number of
segments attached to the membrane:

$$N_s = \frac{N_p}{2} = 2 \pi r_h^2$$

The number of string segments puncturing a concentric spherical
shell with radius $r$ and radial thickness $dr$ is given by

\begin{equation}
dN_p = 8 \pi r dr
\end{equation}

We would like to derive a relation between the string length and
its tension. In general relativity length measurements are
observer-dependent. In the holostar space-time there are two
natural ways to measure the length of a string. An asymptotic
observer at infinity will determine the string length by measuring
(or calculating) the time of flight of a photon travelling along
the full length of the string. A geodesically moving observer in
the holostar's interior space-time, however, will find it more
natural to measure the string length by determining the proper
time it takes himself to travel along the full length of the
string.

\subsection{Point of view of an asymptotic observer at spatial infinity}

Let us first discuss the viewpoint of the asymptotic observer. The
local speed of light in the radial direction in the holostar's
interior, measured by an observer at rest to the coordinate
system, can easily be read off from the metric. It is given by:

\begin{equation}
c_r = \frac{r_0}{r}
\end{equation}

With $dl = dr / c_r$ the length of a string segment $L_p$, ranging
from radial coordinate position $r$ to the boundary membrane then
is given by:

\begin{equation}
L_p(r) = \int_r^{r_h}{dl} = \int_r^{r_h}{\frac{r}{r_0} dr} =
\frac{r_h^2 - r^2}{2 r_0}
\end{equation}

The total length of all string-segments is given by integrating
over all string segments $dN_p$

\begin{equation}
L_{tot} = \int_0^{r_h}{L_p \, dN_p} = \frac{\pi r_h^4}{r_0}
\end{equation}

With $N_s = N_p /2$ the mean string length follows

\begin{equation}
\overline{L} = \frac{2 L_{tot}}{N_p} = \frac{r_h^2}{2 r_0}
\end{equation}

We see that the mean string length, as measured by an observer at
infinity, is inverse proportional to the local value of the string
tension at the membrane. The product of string tension and average
string length is constant and given by:

\begin{equation}
\mu \overline{L} = \frac{1}{16 \pi r_0}
\end{equation}

This value is equal to the pressure of the membrane of a zero
mass-holostar with $r_+ = 0$ and $r_h = r_0$.

It is also possible to calculate the energy of a string segment.
For any one string segment, its total energy is nothing else than
the integral over $dE = \mu dl$. An asymptotic observer at
infinity will calculate the energy to be

\begin{equation}
E_p(r) = \int_r^{r_h}{\mu dl} = \frac{1}{8 \pi r_0}
\ln{(\frac{r_h}{r})}
\end{equation}

The total energy is given by an integral over all string segments
$dN_p$:

\begin{equation}
E_{tot} = \frac{r_h^2}{4 r_0}
\end{equation}

so that the mean energy per string amounts to

\begin{equation}
\overline{E} = \frac{1}{8 \pi r_0}
\end{equation}

Therefore the mean energy and mean length of the strings in the
holographic solution are related to the (local) value of the
string tension at the boundary membrane in the following way:

\begin{equation}
\frac{\overline{E}}{\overline{L}} = 2 \mu = \frac{1}{A} =
\frac{1}{4 S}
\end{equation}

\subsection{Point of view of the geodesically moving interior observer}

The geodesically moving observer has a different measure of
length: His own proper time $\tau$ of travel along the string
length. It can be shown that $r = \tau$ for a geodesically moving
observer \cite{petri/hol}, so that this observer will see quite a
different picture. For him space-time has the appearance of being
flat. The string length, measured in units of proper time of
travel, is nothing else than

\begin{equation}
L_p(r) = r_h-r
\end{equation}

so that the total length of all string-segments amounts to nothing
else than the volume of a sphere in flat 3D-space:

\begin{equation}
L_{tot} = \int_0^{r_h}{L_p \, dN_p} = \frac{4 \pi }{3} r_h^3
\end{equation}

The mean string length then is proportional to the holostar's
gravitational radius:

\begin{equation}
\overline{L} = \frac{2 r_h}{3}
\end{equation}

The mean string length, as measured by a geodesically moving
observer, is inverse proportional to the pressure in the boundary
membrane, whereas the (local) string tension at the boundary
membrane is inverse proportional to the {\em square} of the string
length (measured by the geodesically moving observer).

The mean energy residing in a string segment is calculated by the
geodesically moving observer in the same way as the asymptotic
observer, as an integral of the tension over the whole string
length, with $dE = \mu dl$. A geodesically moving observer moves
nearly radially. Due to the radial boost-invariance of the
holostar space-time the geodesically moving observer measures the
same string tension as the stationary observer. We find

\begin{equation}
E_p(r) = \int_r^{r_h}{\mu dl} = \frac{1}{8 \pi r} \left(1 -
\frac{r}{r_h} \right)
\end{equation}

The total energy again follows from an integral over all string
segments $dN_p$:

\begin{equation}
E_{tot} = \frac{r_h}{2} \simeq M
\end{equation}

We get the remarkable result, that the total energy residing in
all of the strings, as measured by a geodesically moving observer,
is nothing else than the gravitating mass of the holostar.

The mean energy per string then amounts to

\begin{equation}
\overline{E} = \frac{1}{4 \pi r_h}
\end{equation}

which is one fourth of the pressure in the boundary membrane. For
the geodesically moving observer the mean string energy and the
mean string length are related to the string tension at the
boundary by

\begin{equation}
\frac{\overline{E}}{\overline{L}} = 3 \mu
\end{equation}

\section{\label{sec:mach}A coordinate system of strings and Mach's principle}

The radially outlayed strings define a more or less rigid
coordinate system within the whole holostar's interior. If we are
far away from the center, this "coordinate system" is nearly flat.
This has to do with the fact, that the radial metric coefficient
$g_{rr} = r / r_0$ becomes very large at appreciable distances
from the center. Consider an observer at radial coordinate
position $r$ far away from the center. Any {\em proper} sphere
with the observer at its center appears extremely flattened in the
radial coordinate direction. Take the observer's radial coordinate
position to be $r \approx 10^{61}$, corresponding to the current
Hubble-radius of the universe (in natural units). Place a sphere
with proper radius $r_p \approx 10^{61}$ around this observer,
i.e. $r = r_p$. Due to the immense shrinkage of radial ruler
distances, this sphere covers a radial {\em coordinate} interval
range $\delta r = r_p / \sqrt{g_{rr}} \approx 10^{30}$. This is a
factor of $10^{30}$ smaller than $r_p$. Instead of reaching back
to $r-r_p = 0$ the sphere only reaches back to radial coordinate
position $r - \delta r = 10^{60} - 10^{30} = 10^{60}
(1-10^{-30})$.

The interior radial metric coefficient $g_{rr} = r/r_0$ induces an
enormous shrinking of radial ruler-distances. Viewed in the
stationary ($t, r, \theta, \varphi$) coordinate system a proper
sphere whose origin is situated far away from the center is an
extremely "thin", almost membrane-like structure. For a
geodesically moving observer the sphere is even "thinner", due to
Lorentz contraction in the radial direction \cite{petri/hol}. The
strings passing through this proper sphere are parallel to each
other for all practical purposes. The total number of string
segments passing through any such sphere is equal to its
cross-sectional area in the direction perpendicular to $\partial
r$

\begin{equation}
N = \pi r_p^2 = \frac{A_p}{4}
\end{equation}

where $A_p$ is the proper area of the sphere's boundary, measured
in Planck units. This result is independent from the position of
the observer (there might be a small correction for $r \approx
r_{Pl}$). Therefore this result holds {\em locally} within any {\em
arbitrary} space-time region of the holostar's interior. We find
the remarkable result, that the number of string segments (i.e.
the number of fundamental degrees of freedom) in any interior
spherically symmetric region of the holographic solution is
exactly equal to the Hawking entropy.

Note also, that the matter-density (as well as the string tension)
within any sphere with proper radius comparable (or smaller) to
the radial coordinate position of it's center ($r_p <\approx r$)
is nearly uniform with a deviation from homogeneity of the order
$1/ \sqrt{r}$. For a proper sphere with radius equal to today's
Hubble-length ($r_p \approx 10^{61} \, r_{Pl}$), situated at
radial coordinate position $r = 10^{61} \, r_{Pl}$ the
matter-density differs at most by $1 \pm 10^{-30}$. This makes it
clear, that at large distances from the center the holographic
solution is indistinguishable from a homogeneous FRW-model for all
practical purposes.

The coordinate system provided by the strings consists out of real
matter, so we are led to a very Machian viewpoint. The Newton
bucket finally knows - locally ! - why it must accelerate with
respect to the global frame produced by all of the other matter
within the universe. Rotation against the string-frame is locally
detectable, irrespective of the relative alignment of the rotation
axis. Inertial motion with respect to the string frame, however,
will only be detectable in the direction perpendicular to the
strings, due to the boost invariance of any sufficiently small
local frame in the string's longitudinal direction. Any
perpendicular component of the motion is expected to produce an
anisotropy in the frame of an observer moving with nearly constant
velocity, which should be measureable in principle. In fact, such
anisotropies have already been detected \cite{Ralston}, although
their interpretation stands out.

Although the coordinate system provided by the strings is real,
one must keep in mind that the string tension/energy is so low,
and the strings are packed so densely, that we will not be able to
detect their presence directly. The active gravitational
mass-density of a string is zero, so we cannot detect a string by
it's direct gravitational acceleration. There is none. What one
can observe - in principle - is the deficit angle induced by the
strings with respect to a flat geometry. The deficit angle
produces "tidal forces", which can be observed in principle.
However, according to equation (\ref{eq:deficit:angle}) the
deficit angle for a single string is $\Delta \varphi \approx
10^{-122}$ rad at our current position $r \approx 10^{61} r_{Pl}$.
Such a small deficit angle is not observable, neither for a single
string nor any extended space-time region accessible to direct
measurements, such as the solar system. The tidal action of the
strings manifests itself only in the very large scale structure of
the universe, approaching the local Hubble-radius.

\section{Does the cold dark matter consist of low energy strings?}

The holostar solution has been shown in \cite{petri/hol} to be an
astoundingly accurate model for the universe, as we see it today.
For the further discussion I will assume that the holostar
solution actually is the essentially correct description for the
universe.\footnote{Naturally, this is just an assumption. Compared
to the intense study of the FRW-type solutions the properties of
the holographic solution are not very well known. Yet the
theoretical and observational evidence accumulated so far
justifies the assertion, that the holographic solution has a
fairly high {\em potential} to explain many of the phenomena that
are unexplained in the standard FRW-model. Whether it - or a
generalization thereof - will eventually explain {\em all} the
phenomena is an open question. As for any other solution of the
Einstein field equations, it will have to be the tedious task of
comparing theoretical predictions with the vast amount of
observational data, that must guide us to select the solution,
that nature has chosen from the various theoretically possible
choices. The holostar solution is {\em one} such choice. So far it
faired well. Yet it is waiting to be falsified. This should not be
difficult, as it has practically no free parameters that could be
adapted to observation and it does not provide many handles for
modification. One can attempt generalize the solution to the
charged and/or rotating case. These generalizations most likely
will not significantly change the general picture. The charged
holostar solution discussed in \cite{petri/charge} has the same
total interior matter state as the uncharged solution. The only
difference is, that part of the interior mass-energy is of
electro-magnetic origin.}

If this is the case, the form of the stress-energy tensor of the
holostar solution suggests that the universe might actually be a
string dominated structure, in the sense that the {\em dominant}
type of matter resides in strings whereas the "ordinary" matter
(in form of particles) is just a correction/perturbation.

This expectation is quite in agreement with the dynamical
mass-estimates from astronomical observations, which seem to
imply, that a large fraction of matter in our universe is not in
the form of baryonic (ordinary) matter, but resides in a so called
Cold Dark Matter component (CDM). The currently favored FRW-type
models quite clearly require CDM in order to explain the
observational facts. This does not necessarily mean, that CDM must
exist: The experimental determination of the CDM-contribution to
the total energy-density is {\em model-dependent}. A different
model for the universe, such as the holographic solution, might
require quite a different fraction of CDM in order to reconcile
the model with the observations, possibly even no CDM at all. Yet
if the existence of non-baryonic matter turns out to be a real
phenomenon, it seems reasonable to assume that the CDM might be
nothing else than low-energy strings.

Can we estimate the proportion of "stringy matter" with respect to
the "normal" matter (=particles) in the holostar model of the universe?

The starting point for this estimation will be, that at very high
energies, i.e. at the string scale, strings and particles should
be thermalized. The energy density of a string degree of freedom
will be comparable to the energy density of a particle degree of
freedom. If we can determine the fundamental ratio of string to
particle degrees of freedom in thermal equilibrium, we at least
know the (approximate) ratio of the energy densities at the string
scale. Our next task is to estimate how this ratio evolves to the
low energy, low density region of the universe, as we see it
today.

At high temperatures the ratios of the respective degrees of
freedom of strings to particles can be calculated quite easily.
There is one catch: This ratio will be calculated in the context
of general relativity (GR). Although the properties of the
holostar solution suggest that GR is a remarkably
accurate description to the phenomena, even at very high energies,
GR is expected to break down at the string scale.
Whether this break-down will be rather moderate or catastrophic is
not clear at our current state of knowledge. The implicit
assumption in the following derivation is, that GR will only
suffer a moderate break-down. If this is actually the case, we can
interpret the numerical figures derived later as a fairly reliable
order of magnitude estimate.

From the argument given in section \ref{sec:strings} we know, that
there are $N_p = A$ string segments attached to the membrane of
any sufficiently large holostar, where $A$ is the membrane's area.
However, the total number of strings within the holostar is just
half this number, as explained in section \ref{sec:tension}, at
least as long as the number of closed interior loops is
small.\footnote{There is some reason to believe, that the number
of interior closed loops is in fact small. Any interior closed
loop will tend to shrink to its smallest possible size. This
suggests, that an interior closed loop will represent particles,
which justifies the neglection of closed loops: When we compare
the number of degrees of freedom of strings to particles, we
should not count the particle-degrees of freedom when we determine
the string-degrees of freedom and vice versa. What we are actually
comparing, is the number of degrees of freedom of a holostar
consisting {\em exclusively} out of strings to the number of
degrees of freedom of a holostar consisting {\em exclusively} out
of particles. Furthermore, we will see shortly that the number of
particles $N$ with respect to the number of strings $N_s$ in the
holostar solution is small, $N / N_s \approx 1/6$.} See also
Figure \ref{fig:strings}.

In \cite{petri/thermo} the number of particles in thermal
equilibrium at ultra-relativistic energies in a holostar has been
calculated to be $N = A / (4 \sigma)$, where $\sigma$ is the
entropy per particle. The exact value of $\sigma$ depends on the
specifics of the thermodynamic model. The main parameter of the
model is the ratio of fermionic to bosonic (particle) degrees of
freedom. The dependence of $\sigma$ on this ratio is very
moderate. For all practical purposes $\sigma$ lies in the range
$3.15 - 3.3$, which is quite close to the entropy per boson
($\sigma \approx 3.6$) or the entropy per fermion ($\sigma = 4.2$)
of an ultra-relativistic gas with zero chemical potential.

The ratio of the number of strings with respect to the number of
particles in the holostar-solution then is given by:

\begin{equation}
\kappa = \frac{N_s}{N} = 2 \sigma
\end{equation}

Now we proceed to the second task.

It has been shown in \cite{petri/hol}, that the motion of
particles within the interior holostar space-time conserves the
ratio of the energy-densities of different particle species. This
is true for geodesically moving massless and massive particles, as
well as for massive particles following an arbitrary trajectory.
The energy (and entropy-) densities of the different particle
species in the holostar's interior space-time evolve proportional
to $1/r^2 \propto 1/t^2$, irrespective of particle-type.

This is a theoretical result. However, there is some observational
evidence that this characteristic feature of the holostar solution
actually might hold in our universe, at least for {\em
fundamental} particles: The energy densities of electrons and
photons are nearly equal in our universe. See \cite{petri/hol} for
a more detailed discussion. There is even some evidence, that this
assumption might - approximately - hold for compound particles,
such as baryons.\footnote{A nice feature of the static holostar
solution is, that one determine the total number of
ultra-relativistic particle degrees of freedom $f$ at ultra-high
temperatures, when all particles are relativistic, experimentally
and theoretically. This has been done in \cite{petri/thermo}. A
lower bound for $f$ can be derived from the observational data
(total matter-density of the universe, CMBR-temperature),
according to which $f >\approx 6350$. The theoretical result is $f
= 2^4 3^2 4 \pi r_0^2$, where $r_0$ must be evaluated at the
temperature in question. There is some evidence that $r_0^2
\approx 12 / \pi \approx 4$ at the Planck-energy, so that the
theoretical value amounts to $f \approx 7250$. If one assumes that
the ratios of the energy-densities of {\em all fundamental}
particle species in the the holostar space-time with respect to
each other are conserved, one must regard the proton, as the
lightest compound particle, as a "repository" for the frozen out
degrees of freedom at the Planck scale. An electron has four
degrees of freedom, so the proton to electron mass ratio must be
roughly given by $f/4$. Using $f \approx 7250$ one comes quite
close to the true value: $f/4 \approx 1810$, whereas $m_p / m_e
\simeq 1836$.}

What is the case with stringy matter? Here the answer is trivial:
The interior stress energy tensor of the holostar solution is that
of an ensemble of strings. The energy density of the strings at
any radial position is nothing else than the quantity in the $00$
slot of the stress-energy tensor, i.e. $\rho_s = 1/(8 \pi r^2)$.
Therefore, for "stringy matter" $\rho_s \propto 1/r^2 \propto 1/t^2$
holds as well.

Based on these theoretical and observational insights let us make
the general assumption, that the expansion of the universe - as
described by the holostar model - conserves the relative energy-
and entropy densities of the different species of matter.

We are almost ready. We have determined the ratio of the energy
densities of strings and particles at high temperatures, and by
the above proposal this ratio should also be - approximately - the ratio of
the energy densities at any energy.

For a numerical prediction we need the entropy per particle
$\sigma$. At very high energies, i.e. where both particles and
strings are thermalized, it is quite likely that we have a phase
with unbroken supersymmetry. Therefore it seems most appropriate
to take $\sigma = 3.23$, which is the (mean) entropy per
ultra-relativistic particle in a holostar which consists out of
equal numbers of fermionic and bosonic degrees of freedom. With
this figure the ratio of stringy matter to baryonic matter turns
out as:

\begin{equation} \label{eq:kappa}
\kappa = \frac{\rho_s}{\rho_b} = 6.45
\end{equation}

This ratio is quite close to the ratio of CDM to baryonic matter
determined by WMAP, according to which $\kappa \approx 6$.

The crucial assumption that the ratio of the energy densities of
different species is conserved throughout the expansion might hold
only approximately. The various phase transitions that occurred
during the expansion of the universe from the Planck scale to the
low energy scale today might modify this assumption. Whereas there
is some sound theoretical as well as observational evidence, that
this assumption is true when the chemically decoupled particle
species move geodesically (i.e. below the electron-positron
mass-threshold), one cannot expect a priori that the ratios are
unaffected during the complicated phase transitions that took
place at early times, such as the transition from the quark-gluon
plasma to the hadronic phase.

One must also keep in mind, that the WMAP-determination of
$\kappa$ is model dependent, and we are talking here about two
very different models for the universe. One should therefore
compare the ratio in equation (\ref{eq:kappa}) to some other
estimates of the fraction of cold dark matter to baryonic (or
rather "luminous") matter, which are more robust. The analysis of
the rotation curves of galaxies and clusters of galaxies appear to
give higher values. For example, the dynamical mass-estimates for the
Coma-cluster point to a ratio of $\kappa \approx
15-20$.\footnote{Peacock estimates the mean mass-to light ratio
for baryonic matter $M/L \approx 14/h \approx 20$, in units so
that $M/L =1$ for the sun. The mass to light ratio for the
COMA-cluster has been consistently estimated to be $M/L \approx
300-400$. If we take the matter in the COMA-cluster as
representative for the CDM-fraction, we get roughly 15-20 for the
fraction of CDM-matter to baryonic matter.}

\section{Is the universe constructed hierarchically out of strings and membranes?}

If strings (and boundary membranes) are the basic building blocks
of nature, as string-theory claims - albeit so far not with too
much experimental support - it is natural to assume that all of
the matter we see today should be constructed out of these basic
entities. How does this theoretical expectation compare to the
low-energy world we happen to live in today?

Our current understanding is, that the basic building blocks of
the universe are particles and black holes (in the centers of
galaxies, in quasars or as remnants from super-massive stars).
Point-like particles and black holes, which are vacuum-solutions
of the field equations\footnote{The classical black hole solutions
have vacuum everywhere, except for the "matter" that must be
attributed to the point- or ring-like central singularities.}
don't very much look like they could be composed out of strings
and membranes.

The claim, that the universe itself might be nothing else than a
very large holostar appears even more preposterous: The holostar has a
{\em center}. In contrast, the current preferred model of a homogeneous
and isotropic Friedman Robertson-Walker (FRW) universe assumes
from the start, that there is no preferred point in space (however
a preferred time!).

This assumption about how the universe ought to be is called the
cosmological principle. It has guided us quite well, so far. Yet
it is important to remember, that the cosmological principle is
{\em not} a law of nature, but just a convenient assumption, which
allows us to explain the phenomena by a solution of the field
equations with fairly moderate mathematical complexity.
Furthermore the cosmological principle, taken seriously, forces us
to cope with some nasty problems, such as how to explain the
remarkable homogeneity and isotropy of the CMBR. This particular
problem is known as the "horizon problem". One of its solutions is
inflation. But if we truly believe in inflation, the universe {\em
as a whole} is chaotic. We just happen to live in one of its
inflated subcompartments. Depending on the initial conditions, the
primordial chaos will slip in, sooner or later.\footnote{The
cosmological constant and/or the "big rip" might force us to
modify this statement.} Therefore if we are honest, the very idea
that was devised so "save" the cosmological principle at the same
time signals its downfall.

Is then the proposal of a hierarchically constructed universe a
madman's idea which goes against all experience and common sense?
I believe not so. Black holes, the universe and maybe even
particles can be explained quite consistently in terms of strings
and membranes. A successful model at the classical level is the
holographic solution.

\subsection{Black holes}

As should have become clear from section \ref{sec:holo} of this
paper, the holographic solution suggests, that a large black hole
type object can be constructed simply by laying out strings
attached to a spherically symmetric boundary membrane in an
overall spherically symmetric pattern. Although the object
resulting from this construction is not a black hole, in the sense
that it doesn't contain an event horizon, it retains all essential
features of a black hole, most notably it's Hawking temperature
and entropy. This has been shown in great detail in
\cite{petri/thermo, petri/hol}. Yet the main results can be
derived quite effortlessly from the presentation given in this
paper:

The number of strings within the holostar's interior has been
shown to be equal to $N_s = A/2$. For any local observer far away
from the centr we even have $N_s = A/4$, according to the
discussion in section \ref{sec:mach}. The strings quite evidently
constitute the fundamental degrees of freedom of the holostar
solution. It is well known, that the entropy of any large
macroscopic system is proportional to its number of fundamental
degrees of freedom. Therefore the holostar solution predicts $S
\propto A$ (in Planck units) with a factor of proportionality of
order unity. An entropy proportional to area implies a temperature
at infinity $T_\infty \propto 1/M$, i.e. a temperature
proportional to the Hawking temperature, via the thermodynamic
relation $1/T = \partial S /
\partial E$, and using the fact that the total energy $E$ of the
holostar measured at infinity is equal to its gravitating mass
$M$. Now $A \propto r_h^2 \propto M^2$ so that $\partial S /
\partial E \propto M$. This argument demonstrates, that the
holostar is compatible with Hawking's results for black holes, at
least up to a possibly different constant factor.

Furthermore the holostar's boundary membrane, whose properties are
exactly equal to the - fictitious - membrane attributed to a black
hole via the membrane paradigm, guarantees that the holostar's
dynamical action on the exterior space time is practically
equivalent to that of a classical black hole: We know from the
membrane paradigm that all (exterior) properties of a black hole
can be described in terms of its fictitious membrane
\cite{Thorne/mem, Price/Thorne/mem}.

Therefore the assumption, that the black hole type objects in our
universe are rather singularity free holostars, built out of
strings and membranes, is a viable alternative to the black hole
solutions, which are built out vacuum, principally non-localizable
event-horizons and singularities.

Furthermore, it is well known that the Hawking entropy can be
derived rigorously in the context string theory \cite{Strominger/Vafa},
although so far only for extreme or near extreme black holes.
Whereas the string-origin of the Hawking entropy should be more
than obvious from this elegant derivation, the classical vacuum
black hole solutions don't have anything in common with
strings.\footnote{Some authors have interpreted the
ring-singularity in the Kerr-black hole in terms of strings. See
\cite{Burinskii} for such an attempt.} The string nature of the
holographic solution is manifest. One might wonder what route main
stream physics would have taken, if the Hawking entropy-area law
had been first derived in the context of string theory and the
holographic solution had been known at that time.

\subsection{The universe}

What is with the universe itself? In \cite{petri/hol} a fair
amount of evidence has been compiled demonstrating quite clearly,
that the holographic solution - or an extension thereof - might
develop into an alternative model for the universe. Some of the
evidence was summarized in the introduction of this paper. The
evidence is not yet conclusive. Yet there is reason to be
optimistic: The holographic solution has practically no tunable
parameters. It is easily falsifiable. Despite it's "rigid"
structure all of the predictions derived so far were verified
observationally within the experimental errors. Furthermore, the
holographic solution would answer a lot of unanswered questions,
such as the origin of the matter-antimatter asymmetry in curved
space-times \cite{petri/thermo, petri/asym}, the horizon problem
of the standard cosmological models \cite{petri/hol}, the
cosmological constant problem, the problem of singularities,
information-loss, unitary vs. non-unitary evolution, just to name
a few.

If the holographic solution, or an extension thereof, actually
turns out to be a realistic model of the universe, our program to
construct the universe hierarchically out of it's basic building
blocks, strings and membranes, is almost complete.

\subsection{Particles}

One question remains: How do "point-like" particles fit
into this hierarchical picture? All of the fundamental particles
of the Standard Model, i.e. the three generations of quarks and
leptons, appear to be point-like up to the highest energy scales.
There is no experimental evidence yet for any sub-structure.

On the other hand, point-like particles are the cause for severe
problems, already on the purely classical level (see for example
the infinite self-energy of any point-like classical particle).
Some of the difficulties can be circumvented in quantum field
theories. But although renormalization techniques are powerful
tools to control most - and in some situations all - of the
infinities, it is not yet clear, whether a unified description for
{\em all} the phenomena can be devised, that incorporates
point-like particles as one of it's basic building blocks. String
theory suggests otherwise.

The holographic solution itself suggests that elementary particles
might be extended objects of roughly Planck size: The smallest
conceivable holostar has its membrane situated at $r_h = r_0$. It
has a finite boundary area $A_0 = 4 \pi r_0^2$. It is easy to see
from equation (\ref{eq:metric}) that such an "elementary" holostar
has zero gravitational mass: The exterior space-time of such an
object is flat Minkowski space. Elementary particles are
characterized by extremely small masses in natural units. For
example, the proton's mass is $m_p \approx 10^{-19}$ in units of
the Planck mass. Therefore, as a first approximation the masses of
elementary particles can be considered to be zero. Although it is
quite clear, that a spherically symmetric, uncharged holostar
solution cannot describe any realistic particle with non-zero spin
and charge, the solution suggests quite strongly - and quite in
agreement with string theory - that we can have {\em extended}
particle-type objects with masses comparable to the extremely low
masses of elementary particles: The "elementary" zero mass
holostar has a boundary area comparable to the Planck area.

Therefore it is suggestive to interpret "particles" in the
hierarchical picture that has emerged from the previous discussion
not as point-like, but as a spatially bounded collection of
strings and membranes.

Due to the small transverse extension of the strings the interior
"structure" of a particle composed out of strings will only become
apparent at energies approaching the Planck energy. At energies
well below the Planck energy such a particle can be treated as
point-like for all practical purposes.

In the Appendix to this paper a toy-model which "constructs" the
Standard Model of particle physics out of strings and membranes is
given. This naive attempt to reduce the beautiful machinery of the
full 9+1D string theory (or 10+1D M-theory) to the drawing of
suggestive pictures in three spatial dimensions, quite curiously
does a good job in "explaining" some characteristic features of
the Standard Model of Particle Physics. It might prove useful to
promote the imagination of string theorists to devise a realistic
model.

The true nature of the fundamental particles of the Standard Model
will most likely have to be answered by the yet to be found
unified theory of quantum gravity. String theory appears as the
most promising candidate to achieve the unification of all
"forces" into a self consistent picture. Little experimental
guidance did we have so far in accomplishing this monumental task.
But the situation might have changed. We now have a solution to
the Einstein field equations with zero (!) cosmological constant,
which is constructed out of strings and membranes and at the same
time appears to describe the low-energy phenomena in the universe,
as we see it today, rather well. This solution - or an extension
thereof - combined with the dualities of string theory, might
eventually turn out to be a better guide to our understanding of
the phenomena at any high or low energy, than we ever had before.

\section{Discussion}

The holographic solution has been interpreted as a model for the
universe. In this picture the universe is nothing else than a
large black hole type object, whose interior matter state is
dominated by strings. The strings are layed out radially. The
transverse extension of the strings in 3D space has been
determined to be exactly one Planck area, regardless of the
holostar's size.

It was shown, that the large scale properties of the universe as
we see it today arise naturally in a string context. The zero
active gravitational mass-density of the strings implies $H t = 1$
and predicts a nearly unaccelerated expansion. Both predictions
are experimentally fulfilled to a rather good accuracy.

The possibility was explored, whether the cold dark matter
observed in the universe might consist out of stringy matter. An
argument was given which allowed us to estimate the ratio of cold
dark matter to baryonic matter. The estimated ratio $\Omega_{CDM}
/ \Omega_b \approx 6.45$ is quite close to the experimental result
$\Omega_{CDM} / \Omega_b \approx 6$ determined by WMAP. The
argument relies on the characteristic property of the interior
holostar space-time, according to which the (local) ratio of the
energy- and entropy densities of the fundamental particle species
remain constant during the expansion. This theoretically derived
property has some experimental support in the observation, that
the energy-densities of photons and electrons in the universe are
nearly equal.

Some arguments were given, that the universe might be constructed
hierarchically out of its most basic building blocks: strings and
membranes. In this hierarchical picture, black holes (or rather
large black hole type objects) are nothing else than a scaled down
version of the holographic solution. The classical holographic
solution clearly demonstrates the inherent string nature of any
large compact self-gravitating object. The entropy area law for
black holes arises naturally from the string nature of the
solution: The number of string segments puncturing the spherical
boundary membrane of the holographic solution is proportional to
the membrane's area. Every string segment occupies a membrane
segment of Planck area. The total number of fundamental degrees of
freedom (the strings) scales with area. In contrast to a black
hole, the holostar has no event horizon. Its singularity free
interior matter state can be interpreted as the densest
spherically symmetric package of strings. This maximally dense
package is the fundamental reason why a holostar does not collapse
to a singularity, regardless of its size, although its spherically
symmetric boundary membrane lies barely two Planck distances
outside of its gravitational radius.

The holographic solution is an exact solution of the Einstein
field equations with zero cosmological constant. It is well known,
that string theory has severe problems with a non-zero (positive)
cosmological constant. The holographic solution suggests, that the
phenomena can be explained in terms of an exact solution of the
field equations with zero cosmological constant, if we take the
string interpretation of the field equations seriously.

%\bibliographystyle{plain}
%\bibliographystyle{h-physrev}
%\bibliographystyle{hplain}
%\bibliography{bh}

\newpage
\appendix

\section{Are particles a collection of strings and membranes? - a toy model}

The purely classical holographic solutions seems to suggest, that
the universe basically consists of a collection of strings and
membranes. In this appendix I explore the question, whether it
might be possible to incorporate the "fundamental" particles of
the Standard Model of particle physics into this hierarchical picture.

Note, that this section can not and should not be considered as
predictive science, not even by it's own author. It is primarily
based on the drawing of suggestive pictures, which attempt to
capture some of the very abstract results of string theory in a
pictorial representation, guided - or rather misguided - by
intuition alone, and not backed by any serious calculation. The
author does know nothing about string theory! This section doesn't
even intend to propose a serious model for a fundamental particle.
The sole purpose of this appendix is to illustrate in the most
handwaving manner possible, that there is a fair chance to
actually find a construction in the fully developed string
formalism in 10 or 11 dimensions, that explains the properties of
the fundamental particles of the Standard model in terms of
strings and membranes.

Having given ample warning, I will proceed to "construct" the
particles of the Standard Model out of one-dimensional strings and
two-dimensional membranes situated in a space of three spatial
dimensions.\footnote{Quite clearly already the starting point of
the construction has nothing to do with string theory, which is a
theory in 10 or 11 space-time dimension.} The starting point to
this construction is the observation, that

\begin{itemize}
\item the basic building blocks of the universe - in 4D space-time
- appear to be strings and 2D-membranes, according to the
string-interpretation of the holostar solution

\item strings are extended objects, whereas particles are confined
to a small volume
\end{itemize}

If we would like to build particles from strings and membranes, we
must arrange the strings in some closed, bounded structure.
Unfortunately, there are unlimited possibilities to produce closed
objects out of strings and membranes. Therefore we have to reduce
the possibilities to a manageable number.

To accomplish this feat, we first observe, that an object with
entropy 1 consists of four membrane segments, each of Planck size,
according to the Hawking entropy formula. It therefore seems
appropriate to construct our "particles" out of basic building blocks,
consisting out of 4 membrane segments combined into one spherical
membrane with 4 string ends attached:

\begin{figure}[h]
%\begin{minipage}[c]
\begin{center}
\includegraphics[width=10cm, bb=130 300 460 580]{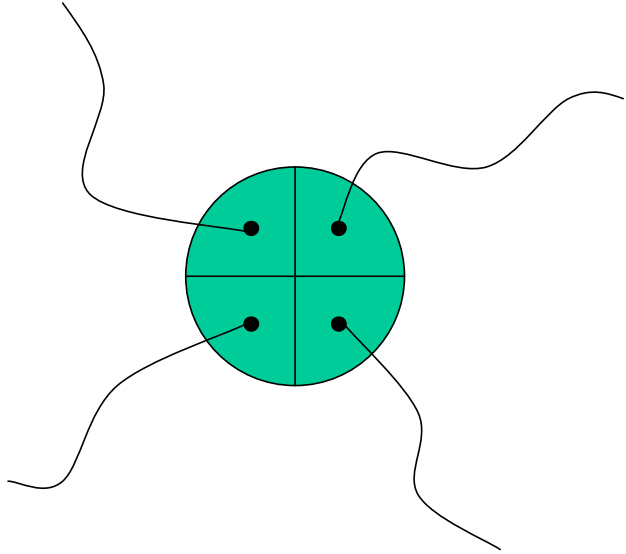}
\caption{\label{fig:s1} Basic building block with entropy 1}
\end{center}
%\end{minipage}
\end{figure}

Let us denote this basic building block with the term s1 (for
entropy = 1).

Second, we wan't to construct the fermions of the Standard Model.
How many basic s1-objects do we require? In \cite{petri/thermo}
the thermodynamics of the holostar's interior matter state was
discussed. If one assumes that the interior matter state consists
of a gas of ultra-relativistic fermions and bosons, the entropy of
an ultra-relativistic fermion turns out somewhat larger than 3,
quite independent from the specifics of the model. This suggests,
that fermionic particles should be constructed out of three
s1-objects (or - more generally - out of 12 membrane segments of
Planck area).

Third, we need a notion of charge. For this we note, that there
are two different possibilities how a string can attach to the
s1-objects. Either the string attaches to two membrane segments on
the {\em same} s1-object, or it forms a link between two {\em
different} s1-objects. In string theory it is known, that charge
can be interpreted as an interactions "within a brane", whereas
gravity is interpreted as interactions "between branes". With this
notion in mind, we label any string attached to the "same brane"
(=s1-object) with a "charge" Q = 1/3.

Now let us construct all possible combinations of strings and
membranes, with the constraint, that every one of the 12 membrane
segments of the three s1-objects has one string end attached.

Curiously, we find four different "particle species". Even more
curiously, these particle species (with our strange "notion" of
charge) correspond exactly to the first generation of particles in
the Standard Model.

We have an "electron". There are three string segments attached to
the same s1-object, which sum up to a total "charge" Q = 1. The
other three string segments form links between different
s1-objects. The construction is symmetric.

\begin{figure}[h]
%\begin{minipage}[c]
\begin{center}
\includegraphics[width=8cm, bb=100 300 460 640]{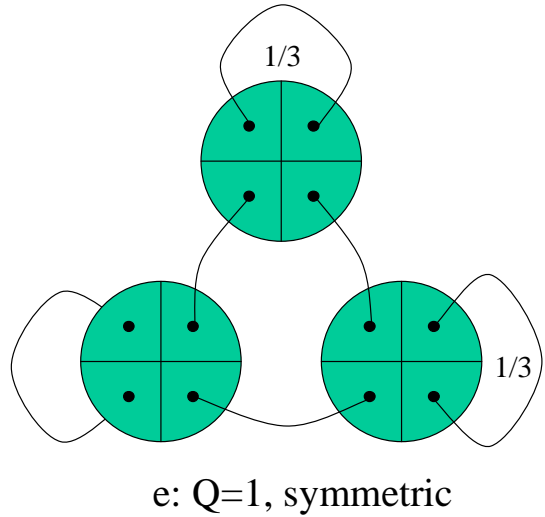}
\caption{\label{fig:electron} electron}
\end{center}
%\end{minipage}
\end{figure}

We have a "neutrino". All six string segments form links between
different s1-objects, there are no "charge loops". The
construction is symmetric.

\begin{figure}[h]
%\begin{minipage}[c]
\begin{center}
\includegraphics[width=7cm, bb=150 280 450 620]{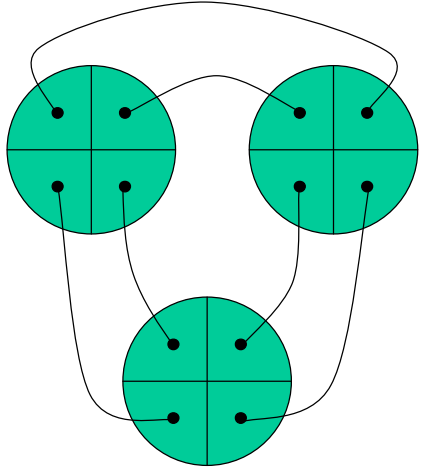}
\caption{\label{fig:neutrino} neutrino}
\end{center}
%\end{minipage}
\end{figure}

We have an (anti-) "down quark". One s1-object contains a "charge
loop". The other five string segments connect to different s1
objects. The total "charge" is Q = 1/3. The construction is not
symmetric. The s1-object that contains the charge can be
distinguished from the other two "uncharged" s1-objects. We
require a symmetry, that symmetrizes (or anti-symmetrizes) the
three possible configurations, i.e. a symmetry that "rotates" the
"charge" between the three s1-objects: "Color"

\begin{figure}[h]
%\begin{minipage}[c]
\begin{center}
\includegraphics[width=7cm, bb=150 230 450 640]{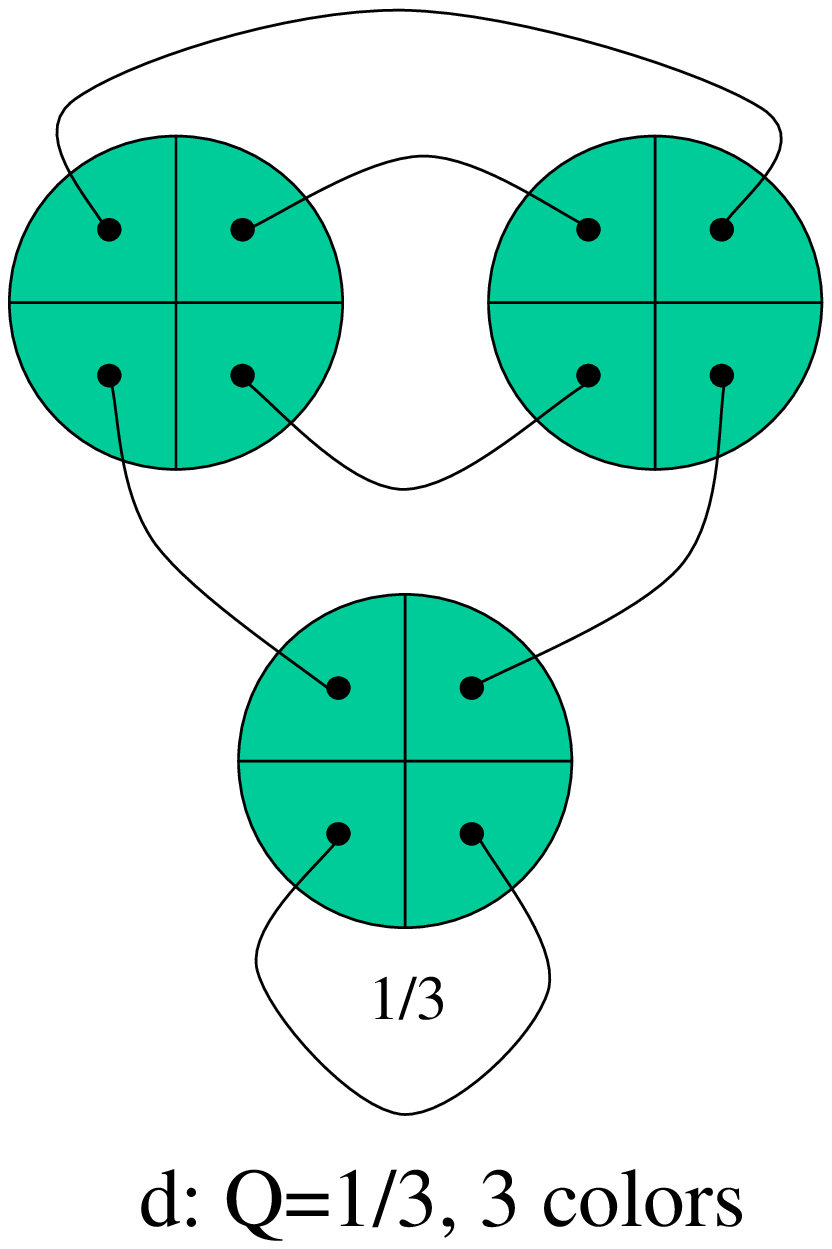}
\caption{\label{fig:down} down quark}
\end{center}
%\end{minipage}
\end{figure}

We have an "up quark": Two of the s1-objects harbor a "charge
loop". The total "charge" is Q = 2/3. Again the construction is
not symmetric and we find "Color".

\begin{figure}[h]
%\begin{minipage}[c]
\begin{center}
\includegraphics[width=10cm, bb=60 330 525 520]{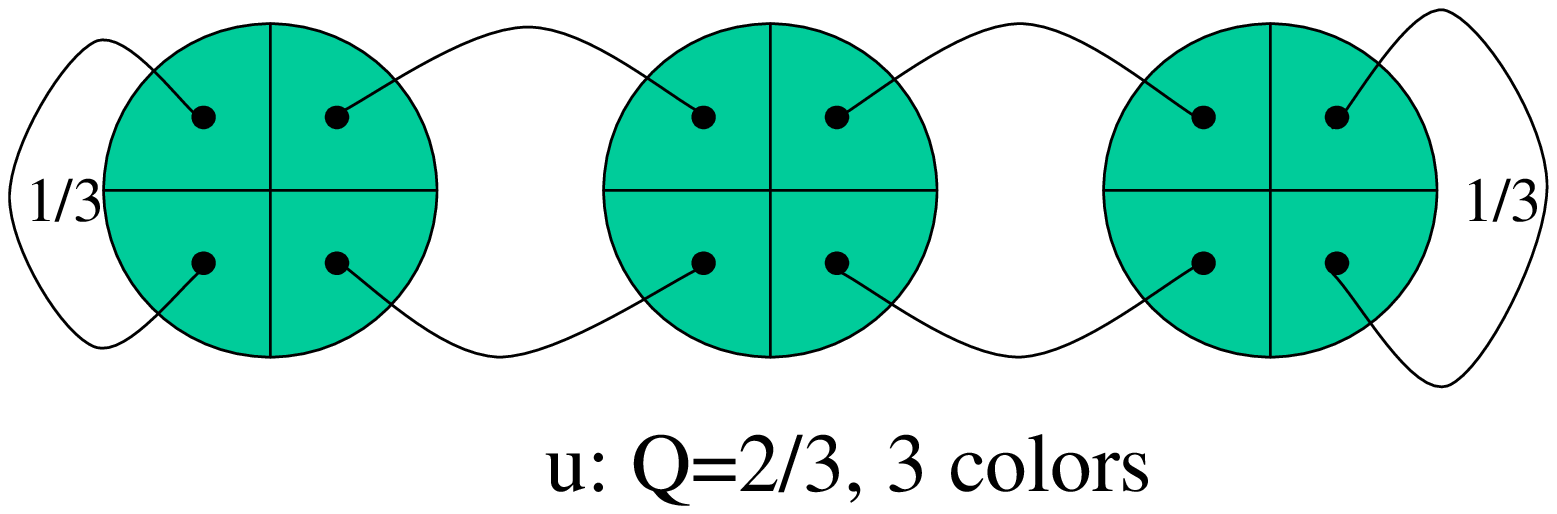}
\caption{\label{fig:up} up quark}
\end{center}
%\end{minipage}
\end{figure}

These are all possible combinations with the rules given
beforehand.

One more handwaving argument suggests itself: As long as the
"particles" are at large distances from each other, each can
preserve it's own identity. The different particle species are
distinguishable. But when the particles come very close to each
other, the individual bonds between the s1-objects will break up,
leaving only the s1-objects. The different fermionic particles are
unified into one description, the GUT scale. If the energy is
turned up even higher, the s1-objects will break up too, so that
there will only be mixture of membranes and strings left, with no
other discernable sub-structure: The string scale. We find that
the pictorial toy-construction makes the "handwaving pictorial
prediction", that the string scale will be higher than the
GUT-scale.

I will leave it as an exercise to the reader, to "prove"
geometrically, that the g-factor of a rotating charged black hole
is exactly 2, under the premise that "charge = strings attaching
to the same membrane" and "mass = strings attaching to a different
membrane". See Misner-Thorne-Wheeler, p. 1149 for the required
geometrical insight to perform this deed.

\end{document}